# Challenging the Long Tail Recommendation

Hongzhi Yin    Bin Cui    Jing Li    Junjie Yao    Chen Chen
Department of Computer Science & Key Lab of
High Confidence Software Technologies
(Ministry of Education), Peking University
{bestzhi, bin.cui, LeaKing, junjie.yao,chenchen}@pku.edu.cn

## ABSTRACT

The success of "infinite-inventory" retailers such as Amazon.com and Netflix has been largely attributed to a "long tail" phenomenon. Although the majority of their inventory is not in high demand, these niche products, unavailable at limited-inventory competitors, generate a significant fraction of total revenue in aggregate. In addition, tail product availability can boost head sales by offering consumers the convenience of "one-stop shopping" for both their mainstream and niche tastes. However, most of existing recommender systems, especially collaborative filter based methods, can not recommend tail products due to the data sparsity issue. It has been widely acknowledged that to recommend popular products is easier yet more trivial while to recommend long tail products adds more novelty yet it is also a more challenging task.

In this paper, we propose a novel suite of graph-based algorithms for the long tail recommendation. We first represent user-item information with undirected edge-weighted graph and investigate the theoretical foundation of applying Hitting Time algorithm for long tail item recommendation. To improve recommendation diversity and accuracy, we extend Hitting Time and propose efficient Absorbing Time algorithm to help users find their favorite long tail items. Finally, we refine the Absorbing Time algorithm and propose two entropy-biased Absorbing Cost algorithms to distinguish the variation on different user-item rating pairs, which further enhances the effectiveness of long tail recommendation. Empirical experiments on two real life datasets show that our proposed algorithms are effective to recommend long tail items and outperform state-of-the-art recommendation techniques.

## 1. INTRODUCTION

Traditionally, most physical retailers concentrate on a relatively small number of established best-sellers, as shown in Figure 1. Economists and business managers often use the Pareto Principle to describe this phenomenon of sales concentration. The Pareto Principle, sometimes called the 80/20 rules, states that a small proportion (e.g., 20%) of products in a market often generate a large proportion (e.g., 80% ) of sales. However, the Internet enjoys the potential to shift this balance. Anderson in his book [3] coined a term-"The Long Tail"- to describe the phenomenon that niche products can grow to become a large share of total sales. In the book, he claimed that Internet technologies have made it easier for consumers to find and buy niche products, which renders a shift from the hit market into the niche market. As shown in Figure 1, the increased popularity of the Internet has accelerated this transition and the Pareto Principle is giving way to the "Long Tail Rule".

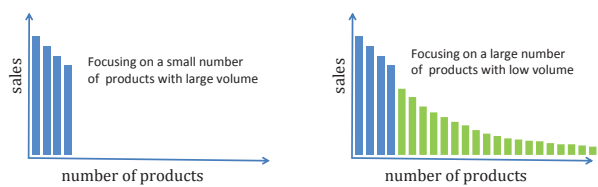

**Figure 1: Hits Market vs. Niche Market**

The long tail brings dual benefits for increasing companies' profit: (1) Compared with popular items, long tail items embrace relatively large marginal profit, which means the endeavor to expand the long tail market can bring much more profit. It is generally accepted in Economics that the economical profit of a completely competitive market is nearly zero. The head market, full of bestselling products is an example of such highly competitive market with little profit. Just as Deniz Oktar, in his recent publication [22], pointed out that for recommender systems, the key to increase profit lies in the exploration of long tail market. He further explained that if a product sells a lot, its profit margin is usually low since all competitors have to sell the same product for the same price; if non-popular products are brought to the interest of right buyers with a successful mechanism, profitability increases drastically. (2) Availabilities to the tail items can also boost the sales on the head due to the so called "one-stop shopping convenience" effect. By providing customers the convenience to obtain both their mainstream and niche goods at one-stop, even small increase in direct revenue from niche products may be associated with much second-order gain due to increased overall consumer satisfaction and resulting repeat patronage. According to results of many analysis [3], companies like Amazon that apply the Long Tail effect successfully make most of their profit not from the best selling products, but from the long tail.

Anderson [3] introduced two imperatives to create a thriving long tail business: (1) Make everything available; (2) Help me find it. The explosion of electronic commerce, such as Amazon.com, Netflix, and the iTunes Music Store, has opened the door to so-called "infinite-inventory" retailers, which makes the former condition fulfilled already by offering an order of magnitude more items than their brick-and-mortar counterparts. But there is still a long way to go for the second imperative since most of the existing recommendation algorithms can only provide popular recommendations, disappointing both to users, who can get bored with the recommender system, and to content providers, who invest in a recommender system for pushing up sales of less known items [6]. In other words, most of the existing recommendation algorithms cannot help users find long tail items with limited historical data due to data sparsity, even if they would be viewed favorably.

For example, association rules have been widely investigated in [24] for item recommendation. The basic idea of mining association rules is simple: If many users rating $item_1$ (high support) also rate $item_2$ (high confidence), then a new user rating $item_1$ is most likely to rate $item_2$ in the near future. Considering that the mined association rules require high support for $item_1$ and $item_2$, they typically recommend rather generic, popular items. Similarly, the recommendation results from the classic collaborative filtering(e.g. [12, 20]) are always local popular and obvious. The basic idea of collaborative filtering is as follows: for a given user, it first finds $k$ most similar users by using Pearson correlation or Cosine similarity to compute similarity among all users, and then recommends the most popular item among these $k$ users [7]. Recently, various latent factor models, such as matrix factorization model [1, 2, 16] and probabilistic topic model [15, 5], have been proposed to make recommendations. Although these models perform well in recommending popular items, they cannot address the problem of long tail item recommendation because these models can only preserve principal components and factors while ignoring those niche latent factors, and these principal components and factors can only capture properties of the popular items [6].

In addition, recent studies [7] show that most of popular recommenders can even lead to a reduction in sales diversity rather than enhancement, but they do not provide any solution to promote sales diversity. Although most of existing recommenders can push each person to new products, they often push different users toward the same products. These recommenders create a rich-get-richer effect for popular products and vice-versa for unpopular ones. So a new recommender model that can both promote sales diversity and help users discover their favorite niche products is most desirable.

In this paper, we investigate the novel problem of long tail recommendation, and propose a suite of recommendation algorithms for this task, including Hitting Time, Absorbing Time and Absorbing Cost algorithms. The proposed methods have several advantages over existing ones. First, our proposed algorithms have the desirable ability to help users accurately find their favorite niche items, rather than merely popular items, which enables less mainstream music or film to find a customer. Second, we treat various user-item rating pairs differently in our algorithms, which drastically improves the accuracy of long tail recommendation. Third, our proposed algorithms can provide more diverse recommendations to different users in the aggregate, which lays solid foundation for promoting sales diversity. In a word, our proposed recommendation algorithms provide an effective novel framework, or an alternative way for building a real recommender system.

To the best of our knowledge, this is the first work proposed to address the long tail recommendation problem, and our work contributes to its advancements in the following ways:

- We analyze the long tail phenomenon and long tail recommendation problem, and propose a basic solution called *Hitting Time* algorithm based on user-item graph. To further improve recommendation diversity and accuracy, we extend the hitting time and propose efficient *Absorbing Time* algorithm.

- We propose *entropy-cost* model to distinguish the variation on different user-item rating pairs, based on which we design *Absorbing Cost* algorithms to improve the recommendation accuracy and quality.

- We propose a new LDA-based method to mine and discover users' latent interests and tastes by using only the rating information, which provides a reasonable and effective way to compute *user entropy*, a novel feature proposed in this work.

- We conduct extensive experiments to evaluate our proposed algorithms, as well as other state-of-the-art recommendation techniques, using two real datasets. The experimental results demonstrate the superiority of our methods in various measurements including recall, diversity and quality for long tail item recommendation.

The rest of the paper is organized as follows. We review the related work in Section 2. In Section 3, we present an overview of the long tail recommendation and propose a basic solution based on hitting time. In Section 4, we propose two novel approaches to improve the basic solution, which enhance both the efficiency and effectiveness of long tail recommendation. In Section 5, we conduct extensive experiments and present an empirical study on two real datasets. Finally, we offer our concluding remarks in the last section.

## 2. RELATED WORK

A major task of the recommender system is to present recommendations to users. The task is usually conducted by first predicting a user' s ratings (or probability of purchasing) for each item and then ranking all items in descending order. There are two major recommendation approaches: content-based filtering and collaborative filtering.

The content-based recommendation [25] is based on the assumption that descriptive features of an item (meta data, words in description, price, tags, etc.) tell much about a user's preferences to the item. Thus a recommender system makes a decision for a user based on the descriptive features of other items the user likes or dislikes. However, in e-commerce systems, products usually have very limited description (title, user reviews, etc.). The effectiveness of content-based approaches is limited.

In the collaborative filtering approach [12, 20], user behavior history is utilized to make recommendations. This approach is based on the assumption that users with similar tastes on some items may also have similar preferences on other items. Thus the main idea is to utilize the behavioral history from other like-minded users to provide the current user with good recommendations. Research on collaborative filtering algorithms has reached a peak due to the 1 million dollar Netflix movie recommendation competition. Factorization-based collaborative filtering approaches, such as the regularized Singular Value Decomposition, perform well on this competition, significantly better than Netflix's own well-tuned Pearson correlation coefficient (nearest neighbors) algorithm. Recently, authors in [6] conduct extensive experiments to evaluate the performances of various matrix factorization-based algorithms and neighborhood models on the task of recommending long tail items, and their experimental results show that: (1) the accuracy of all algorithms decreases when recommending long tail products, as it is more difficult to recommend non-trivial items; (2) PureSVD outperforms other state-of-the-art matrix factorization based algorithms and neighborhood models.

In addition, a group of probabilistic topic models have been developed in recent years. [15, 5] use LDA-based methods for community recommendation. [17] uses an approach based on LDA for

897

recommending tags. Besides, some graph-based recommender systems are proposed. Authors in [29, 23] apply a node similarity algorithm to item-based recommendation. Hitting time and Commute time models have been proposed as a basis for making recommendations [8] in which the target users are set to be starting states instead of absorbing states, but the recommending from their hitting time and commute time is almost the same as recommending the most popular items identified by the stationary distribution, often causing the same popular items to be recommended to every consumer, regardless of individual consumer taste [4].

In contrast to our work, most of existing recommendation algorithms can not recommend long tail items with limited historical data due to data sparsity. Meanwhile, the diversity of their recommendations is poor. To challenge long tail item recommendation and improve recommendation diversity, we propose a novel suite of graph-based algorithms in this paper.

## 3. OVERVIEW OF THE LONG TAIL RECOMMENDATION

In this section, we first introduce the graph modeling of the user-item information, and then propose a basic solution called Hitting Time to help users find their favorite long tail items on this graph.

### 3.1 Graph-based Representation of User-Item information

Users and items in recommendation scenarios are inter-connected through multiple ways, and we will model these information by an edge-weighed undirect graph $G(V, E)$ where $V$ denotes the set of nodes and $E$ represents the set of edges. $n$ is the number of nodes and $m$ is the number of edges. This graph has weights $w(i,j)$: if there is an edge connecting $i$ and $j$, then $w(i,j)$ is positive; otherwise, $w(i,j) = 0$. Since the graph is undirected, the weights are symmetric. The weight $w(i,j)$ indicates the strength of the relation between node $i$ and node $j$. Let $A = (a(i,j))_{i,j \in V}$ be the adjacent matrix of the graph with $a(i,j) = w(i,j)$.

Let us consider a movie-rating dataset. Basically, this dataset consists of three kinds of information; demographic information about the users, information about the movie, and information about rating which users assign to movies they have watched. As shown in Figure 2, each element of the user and movie sets corresponds to a node of the graph, the has_watched link is expressed as an edge, and the edge weight corresponds to the rating. This graph expresses the relation between users and items (a user-item graph). In this graph, nodes can be divided into two disjoint sets such as $V = V_1 \bigcup V_2$, where $V_1$ denotes the set of users and $V_2$ represents the set of movies. A query node corresponds to a user to whom the recommender system wants to make recommendations.

Now we define the Long Tail Recommendation problem as follows:

PROBLEM 1. *Given : a user-item graph $G(V, E)$ with adjacency matrix A, and a query user node q. Find: top-k item nodes that (1) are close enough to q, and (2) lie in the long tail.*

### 3.2 Random Walk Similarity

Since the essence of recommendation task is to find user-interested items which are typically close to user preferences in the graph, random walk can be applied to a user-item graph $G$ specified by the adjacent matrix $A$, where various statistics of this walk are useful for computation of proximity [4]. The Markov chain describing the sequence of nodes visited by a random walker is called random walk. A random walk variable $s(t)$ contains the current state of the

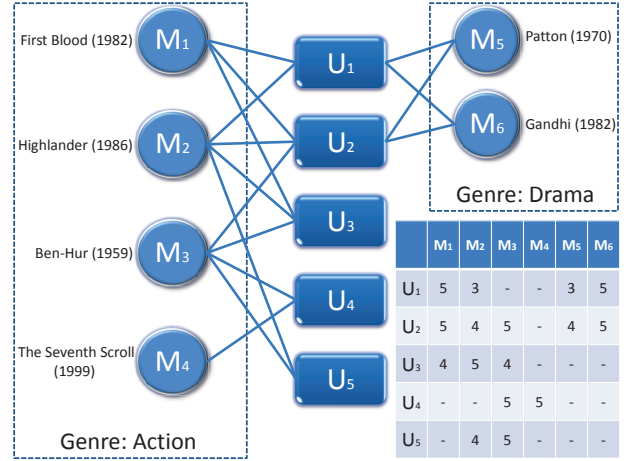

**Figure 2: Example of a user-item bipartite graph**

Markov chain at time step $t$: if the random walker is in state $i$ at time $t$, then $s(t) = i$. The random walk is defined with the following single-step transition probabilities of jumping from any state or node $i = s(t)$ to an adjacent node $j = s(t+1)$ as follows:

$$p_{ij} = P(s(t+1) = j | s(t) = i) = \frac{a(i,j)}{d_i} \quad (1)$$

where $d_i = \sum_{j=1}^{n} a(i,j)$, denotes the degree of node $i$.

Since the graph is connected, the Markov chain (first-order) is irreducible, that is, every state can be reached from any other state. If we denote the probability of being in a state $i$ at time $t$ by $\pi_i(t) = P(s(t) = i)$ and we define $\mathbf{P}$ as the transition matrix with entries $p_{ij} = P(s(t+1) = j | s(t) = i)$, the evolution of the Markov chain is characterized by $\pi(t+1) = \mathbf{P}^T \pi(t)$ where T is the matrix transpose and $\pi(t) = [\pi_1(t), ..., \pi_i(t), ..., \pi_n(t)]^T$. This provides the state probability distribution $\pi(t)$ at time $t$ once the initial probability distribution $\pi(0)$ is known. It is well-known [27] that such a Markov chain of random walk on a graph has the stationary probabilities as follows:

$$\pi_i = \frac{\sum_{j=1}^{n} a(i,j)}{\sum_{i,j=1}^{n} a(i,j)} \quad (2)$$

which means, the stationary probability $\pi_i$ is proportional to $d_i$.

There are some random walk based methods to measure the proximity between a pair of nodes ¡i, j¿ in recommendation: random walk with restart [23], commute time [4, 8], Katz [8], Random forest based algorithm(RFA) [8], just to name a few. Although these methods can give reasonable proximity scores, they can not challenge long tail item recommendation. Some of them do not take into account the popularity of items, e.g., Katz and RFA; others such as random walk with restart and commute time tend to recommend popular items. Their recommendations are very similar to the results by simply suggesting the most popular items, most likely because they are both dominated by the stationary distribution, as stated in [4, 8].

So in the following part, we propose a basic solution called Hitting Time based on the random walk, and investigate its theoretical foundation to recommend long tail items.

### 3.3 Basic Solution Based on Hitting Time

In terms of the stationary distribution mentioned above, it is easy to formulate the property of time-reversibility [18]: it is equivalent to say that for every pair $i, j \in V$, $\pi_i p_{i,j} = \pi_j p_{j,i}$, where $p_{i,j}$ denotes the probability that a random walker starting from node $i$ reaches node $j$. The probability $p_{i,j}$ can capture the graph structure where $i$ and $j$ have many common neighbors and/or many short paths in between them. Specifically, the probability has the nice property of increasing when the number of paths connecting two nodes increases and when the length of any path decreases. Due to this property, the probability $p_{i,j}$ can be used to compute the similarity between $i$ and $j$. In short, two nodes are considered as being similar if there are many short paths connecting them.

From the property of time-reversibility, we can see that the probabilities $p_{i,j}$ and $p_{j,i}$ are not symmetric. So there can be two alternatives to compute the similarity between two nodes on the graph. Given one query user node $q$, we can apply either $p_{q,j}$ or $p_{j,q}$ to compute the similarity between $q$ and $j$ in principle. One alternative is to apply $p_{q,j}$ to find the most relevant node $j$ for the query node $q$ on the graph. But this method can not challenge the long tail recommendation. According to the property of time-reversibility, we can get the following equations:

$$p_{q,j} = \frac{p_{j,q} \pi_j}{\pi_q} \quad (3)$$

where the denominator $\pi_q$ is fixed for the given query node $q$. From the above equation, we can see that the popular item node $j$ will enjoy priority to be suggested by $p_{q,j}$ because $p_{q,j}$ is most likely to be dominated by the stationary distribution $\pi_j$.

In order to recommend long tail items, we adopt $p_{j,q}$ to compute the relevance between $q$ and $j$. Similar to the equation 3, we get the equation for $p_{j,q}$ as follows:

$$p_{j,q} = \frac{p_{q,j} \pi_q}{\pi_j} \quad (4)$$

where $\pi_q$ is fixed for the given query node $q$. It can be seen from the above equation that $p_{j,q}$ represents the ratio between the relevance score $p_{q,j}$ and the stationary probability $\pi_j$. As mentioned in Section 3.2, the stationary probability $\pi_j$ is proportional to $d_j$, the degree of node $j$. So the probability $p_{j,q}$ discounts items by their overall popularity. Based on the probability $p_{j,q}$, we introduce the definition of hitting time from $j$ to $q$, denotes as $H(q|j)$.

DEFINITION 1. *(Hitting Time). The hitting time from $j$ to $q$, denoted as $H(q|j)$, is the expected number of steps that a random walker starting from node $j$ ($j \neq q$) will take to reach node $q$.*

By definition of hitting time, we can easily get the following equation:

$$H(q|j) = \frac{1}{p_{j,q}} = \frac{\pi_j}{p_{q,j} \pi_q} \quad (5)$$

According to the above equation, a small hitting time from an item node $j$ to the query node $q$ means: (1) $q$ and $j$ are relevant (2) the item node $j$ is with low stationary probability, which implies fewer users rate or purchase the item $j$.

Now we transfer the Long Tail Recommendation problem into the following problem :

PROBLEM 2. *Given : a user-item graph $G(V, E)$ with adjacency matrix A, and a query user node q. Find : top-k item nodes with smallest hitting times to q.*

Based on the above analysis, we can use the hitting time to recommend long tail items. Given a query user, we first compute the hitting times from all item nodes to the query user node, and then use this measure to rank all items except those already rated by the query user. Finally, we select $k$ items with smallest hitting times as recommendations to the query user. For example, as shown in Figure 2, $U_5$ is the assumed query user node, and we can compute the hitting times $H(U_5|M_i)$ for all movies except those already rated by $U_5$: $H(U_5|M_4) = 17.7$, $H(U_5|M_1) = 19.6$, $H(U_5|M_5) = 20.2$ and $H(U_5|M_6) = 20.3$. So, we will recommend the niche movie $M_4$ to $U_5$ since it has the smallest hitting time to $U_5$, while traditional CF based algorithms would suggest the local popular movie $M_1$. From the figure, we can see that $M_4$ is not only harmony with the taste of $U_5$ (e.g., Action movies), but also unpopular, rated by only one user.

## 4. ENHANCEMENTS OF RECOMMENDATION

In the last section, we propose the user-based Hitting Time algorithm to challenge long tail item recommendation. It is well known that the accuracy of recommendation methods depends mostly on the ratio between the number of users and items in the system [25]. In large commercial systems like Amazon.com where the number of users is much greater than the number of items, the average number of ratings per item is much higher than the average number of ratings per user, which means every item has more information to use. To utilize more information and improve the accuracy of recommendation, we refine the Long Tail Recommendation problem as follows:

PROBLEM 3. *Given : a user-item graph $G(V, E)$ with adjacency matrix A, and a query user node q. Find : top-k item nodes which (1) are close enough to $S_q$, and (2) lie in the distribution of long tail .*

where $S_q$ denotes the set of items rated(or purchased) by query user $q$. We would like to have a recommendation algorithm for suggesting items which not only relevant to user preferred item set $S_q$, but also hard-to-find.

Based on the above problem definition, we extend the Hitting Time algorithm and propose two efficient and effective algorithms for the long tail recommendation in this section.

### 4.1 Recommender Based on Absorbing Time

In this part, we extend the Hitting Time and propose an efficient item-based algorithm called Absorbing Time to challenge the long tail recommendation. First, we give the definition of Absorbing Nodes and Absorbing Time, as follows.

DEFINITION 2. *(Absorbing Nodes). In a graph G, a set of nodes S are called absorbing nodes if a random walker stops when any node in S is reached for the first time.*

DEFINITION 3. *(Absorbing Time). Given a set of absorbing nodes S in a graph G, the absorbing time denoted as $AT(S|i)$ is the expected number of steps before a random walker, starting from node i, is absorbed by S.*

where $S$ is a subset of $V$. Note that when $S = \{j\}$, the absorbing time $AT(S|i)$ is equivalent to the hitting time $H(j|i)$. Let $s(t)$ denotes the position of the random walk at discrete time $t$. The first-passage time $T^S$ is the first time that the random walk is at a node in $S$. Thus $T^S = \min\{t : s(t) \in S, t \geq 0\}$. It is obvious that $T^S$ is a random variable. The absorbing time $AT(S|i)$ is the expectation of $T^S$ under the condition $s(0) = i$, that is, $AT(S|i) = E[T^S|s(0) = i]$. The absorbing time $AT(S|i)$ denotes the expected number of steps before node $i$ visits the absorbing nodes $S$. The recurrence relations for computing absorbing time can be obtained by first step analysis [14, 13].

$$AT(S|i) = \begin{cases} 0, & i \in S \\ 1 + \sum_{j=1}^{n} p_{ij} AT(S|j), & i \notin S \end{cases} \quad (6)$$

The meaning of the above recurrence formula is : in order to jump from node $i$ to the absorbing nodes $S$, one has to go to any adjacent state $j$ of state $i$ and proceed from there. The absorbing time can also be interpreted as weighted path-length from node $i$ to node $j \in S$ as follows:

$$AT(S|i) = \sum_{j \in S} \sum_{path(i,j)} |path(i,j)| Pr(path(i,j)) \quad (7)$$

where $\sum_{j \in S} \sum_{path(i,j)} Pr(path(i,j)) = 1$, $|path(i,j)|$ is the length of the path from $i$ to $j$ and $Pr(path(i,j))$ is the probability of following the $path(i,j)$ from $i$ to $S$.

Now we transfer the Long Tail Recommendation problem into the following problem :

PROBLEM 4. *Given : a user-item graph $G(V, E)$ with adjacency matrix $A$, and a query user node $q$. Find : top-k item nodes with smallest absorbing time $AT(S_q|j)$ to $S_q$.*

Based on the above definition and analysis, we can use the Absorbing Time to recommend long tail items. Given a query user $q$, we can first compute the absorbing times $AT(S_q|i)$ for all items based on the graph $G$, and then use this measure to rank all items except those in $S_q$. Finally, we select $k$ items with smallest absorbing times as recommendations to user $q$. However, there are still two concerns for using the straightforward absorbing time.

- The graph $G$ can be excessively large (e.g., millions of users and items). In fact, most nodes are irrelevant to the query user, but they increase the computational cost.

- Solving the linear system can be time-consuming. When number of variables of the linear system is millions, it becomes extremely inefficient to get an exact solution to that linear system. The absorbing times, starting from every non-absorbing node and absorbed by $S$ can be computed in time $\mathcal{O}(n^3)$ where $n$ is $|V| - |S|$ [13].

To overcome these two obstacles, we propose the following efficient algorithm for long tail recommendation using Absorbing Time in Algorithm 1. Similar to [10], we compute Absorbing Time by iterating over the dynamic programming step for a fixed number of times $\tau$ rather than directly solving the linear system. This leads to the truncated Absorbing Time, which is reasonable and appropriate in the context of recommendation because what we really care about is the ranking of item nodes rather than the exact value of Absorbing Time. A good selection of $\tau$ would guarantee that the ranking of top $k$ items stays stable in the future iterations. If the algorithm is implemented on the global graph, for each user its complexity is $\mathcal{O}(\tau \cdot m)$, where $m$ denotes the number of edges and $\tau$ denotes the iteration number. In order to scale the search and improve the efficiency, we first deploy a breadth-first search from the absorbing node set and stop expansion when the number of item nodes exceeds a predefined number $\mu$. Then we apply the iterative algorithm to the local subgraph. A worst-case analysis suggests that $m = \mathcal{O}(\mu^2)$ on the subgraph. Thus, the worst-case running time of the Absorbing Time based on the local subgraph is $\mathcal{O}(\tau \cdot \mu^2)$. However, in practice, the rating matrix is very sparse and the running time of the algorithm is much less than this worst-case analysis suggests. In addition, $\mu$ and $\tau$ do not need to be large, which ensures the efficiency of this algorithm. Generally, when we use 15 iterations, it already achieves almost the same results to the exact solution which we can get from solving the linear system.

**Algorithm 1** Recommendation Using Absorbing Time

**Input:**
   A user-item graph, $G(V, E)$ with adjacency matrix $A$;
   A query user node, $q \in V$;

**Output:**
   A ranked list of $k$ recommended items;

1: Given the query user $q$, find the item set $S_q$ rated by $q$.
2: Construct a subgraph centered around $S_q$ by using breadth-first search in the graph $G$. The search stops when the number of item nodes in the subgraph is larger than a predefined number of $\mu$.
3: Initialize the absorbing times of all nodes in the subgraph with $AT_0(S_q|i) = 0$.
4: For all nodes in the subgraph except those in $S_q$, iterate

$$AT_{t+1}(S_q|i) = 1 + \sum_j p_{ij} AT_t(S_q|j)$$

   for a predefined number of $\tau$ iterations.
5: Let $AT_\tau(S_q|i)$ be the final value of $AT_t(S_q|i)$. Output the items $R_q$ which have the top-$k$ smallest $AT_\tau(S_q)$ as recommendations.

## 4.2 Recommender Based on Absorbing Cost

The proposed efficient Absorbing Time algorithm has the desirable ability to help users find their favorite long tail items. However, it lacks modeling of rich information to distinguish the variation on different user-item rating pairs besides the rating scores, thus there is still much space to improve its accuracy for the long tail recommendation. Intuitively, a score rated by a user who specializes in limited interests would provide much more valuable information than the rating offered by a person of wide interests. However, in the Absorbing Time model, those two rating scores are not treated distinguishingly as it should be, resulting in the loss of important information which can otherwise be utilized to reach more accurate recommendation.

As show in Figure 2, $M_3$ is rated by both $U_2$ and $U_4$ with the same rating score 5-stars. There is a critical question when only considering the raw rating score: Should the equal ratings provided by different users to the same item be viewed equally important in the recommendation? Or are the costs of jumping from a certain item node to different connected users with same ratings in the bipartite graph equal? Clearly not. In this case, at an intuitive level, the rating from $U_4$ to $M_3$ may capture more meaningful information, or be more important than the rating from $U_2$ to $M_3$. In other words, jumping from $M_3$ to $U_4$ should be less costly than jumping from $M_3$ to $U_2$. The key difference is that the connected users are different: $U_4$ is a taste-specific user (e.g., only likes Action movies) while $U_2$ has a wider range of interest (e.g., both Action and Drama movies).

To treat various user-item rating pairs differently, we propose the Absorbing Cost model in this part. In a similar way to the Absorbing Time, the Absorbing Cost is the average cost incurred by the random walker starting from state $i$ to reach absorbing states $S$ for the first time. The transition cost from state $i$ to its adjacent state $j$ is given by $c(j|i)$. Notice that the Absorbing Time $AT(S|i)$ is obtained as a special case where $c(j|i) = 1$ for all $i, j$. The following recurrence relations for computing Absorbing Cost $AC(S|i)$ can easily be obtained by first-step analysis [14, 13].

$$AC(S|i) = \begin{cases} 0, & i \in S \\ \sum_j p_{ij} c(j|i) + \sum_j p_{ij} AC(S|j), & i \notin S \end{cases} \quad (8)$$

In the following parts, we first propose a novel feature called *user entropy*, and then utilize this feature to compute the transition cost and capture the variation on different user-item rating pairs.

### 4.2.1 Entropy Metric used in Information Theory

Before performing a theoretical analysis, we first briefly review the entropy in information theory [26]. Suppose there is a user who has a wide range of interests and tastes, which tends to increase the ambiguity of the user. On the other hand, if the user centers on few specific interests, this tends to increase the specificity of the user. Therefore, a single rating from a *specific* user is most likely to be more important to distinguish the specificity of the item than another rating from an *ambiguous* user.

For example, if two different items are connected by a *specific* user, then they are more likely to be similar and proximal in topics and tastes. Based on the intuition, we introduce a novel feature *user entropy* to weight the cost of jumping from one item node to different connected user nodes as shown in Equation 9. Specifically, we currently only consider changing the cost of jumping from the item node to the user node, and keeping the cost of jumping from the user node to the item node as a predefined constant $C$.

$$AC(S|i) = \begin{cases} 0 &, i \in S \\ \sum_j p_{ij} E(j) + \sum_j p_{ij} AC(S|j), & i \in V_2 - S \\ C + \sum_j p_{ij} AC(S|j) &, i \in V_1 - S \end{cases} \quad (9)$$

where $V_1$ is the set of all user nodes and $V_2$ is the set of all item nodes in our system; $E(j)$ denotes the entropy of user $j$ and $C$ is a tuning parameter, which corresponds to the mean cost of jumping from $V_2$ to $V_1$ in our current model.

In the following subsections, we propose two strategies to compute the novel feature *user entropy*. They are item-based and topic-based respectively.

### 4.2.2 Item-based User Entropy

The Absorbing Time algorithm and most CF-based recommendation algorithms only consider the information of users' ratings of items, and treat different user-item rating pairs equally even if some users are very general and prefer everything. More generally, a great variation in users' preferred item distribution is likely to appear, and it may thus cause the loss of important information since different user-item pairs are not sufficiently distinguished. It is easy to understand that the broader a user's tastes and interests are, the more items he/she rates and prefers. Thus general users would have a larger collection distribution than the taste-specific ones, which tends to increase the ambiguity and uncertainty of the users in the ordinary sense.

An assumption is accepted that if a user rates (or downloads, purchases) large number of items, especially with equal probability, the ambiguity (uncertainty) of the user tends to increase, otherwise if a user rated only a few items, the specificity of the user tends to increase. Using information theory, the entropy [26] of a user $u$ is defined as

$$E(u) = -\sum_{i \in S_u} p(i|u) \log p(i|u) \quad (10)$$

where $p(i|u) = \frac{w(u,i)}{\sum_{i \in S_u} w(u,i)}$.

### 4.2.3 Topic-based User Entropy

In the last section, we propose an item-based method to compute *user entropy*. But the assumption made in the item-based method is not always true, because it is possible that an interest-specific user rates a large number of items which all center around his specific interest. So we try to employ probabilistic topic model [28] to directly model users' latent interests and tastes in this section.

Given a topic set $T = \{z_1, z_2, ..., z_K\}$ and a user $u$'s topic distribution $\theta_u$, we can compute his/her entropy as follows:

$$E(u) = -\sum_{z_i \in T} p(z_i|\theta_u) \log p(z_i|\theta_u) \quad (11)$$

To compute a user's latent topic distribution, we propose a new LDA based method to learn users' latent topics and tastes by only utilizing the rating information. In our method, user-item data is entered as a frequency count where the value is the rating score. We choose the rating to measure the strength of relationship between users and items. A user's preference to different items may be various, for example, to his favorite item, the user may give high rating score. We use $w(u, i)$, which is the relation between the user $u$ and his/her rated $i_{th}$ item $item_{u,i}$, to denote $u$'s rating score of $item_{u,i}$. In our method, $w(u, i)$ is viewed as the frequency of the item's appearance in the item set $S_u$ rated by $u$. As shown in Figure 3, we denote the per-user topic distribution as $\theta$, the per-topic item distribution as $\phi$, and the generative process of LDA model can be summarized as follows:

1. For each topic $k = 1, ..., K$, draw a multinomial $\phi_k$ from a Dirichlet prior $\beta$.

2. For each user, $u \in U$:

    (a) Draw $\theta_u \sim Dirichlet(\cdot|\alpha)$.

    (b) For each item, $item_{u,i}$, in user rated item set $S_u$:

        i. Draw $z$ from multinomial $\theta_u$;
        ii. Draw $item_{u,i}$ from multinomial $\phi_z$;
        iii. Repeat the above two steps $w(u, i)$ times.

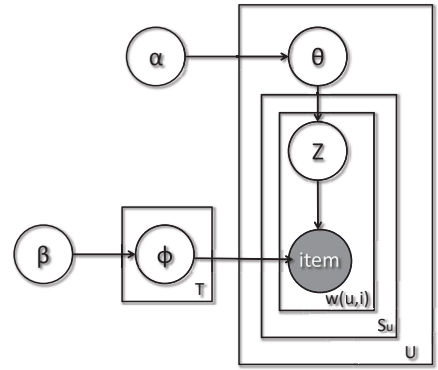

Figure 3: LDA model for user-item data

Below, we show how to train LDA and estimate parameters using Gibbs sampling, then infer the per-user topic distribution. Following the same procedure of assigning each word with different topics in Gibbs sampling [9], we start our algorithm by randomly assigning a topic set $T = \{z_1, z_2, ..., z_{w(i,j)}\}$ to the $j_{th}$ item of the $i_{th}$ user. Figure 4 demonstrates the process. For example, we assign a topic set $T_{ij} = \{z_{i,j,1}, z_{i,j,2}, ..., z_{i,j,w(i,j)}\}$ to $item_{i,j}$. Then each topic in the topic set is updated by Gibbs sampling using:

$$P(z_{i,j,k} = z | T_{-(i,j,k)}, item) \propto \frac{n^{item_{i,j}}_{-(i,j,k),z} + \beta}{n^{\bullet}_{-(i,j,k),z} + N_I\beta} \frac{n^{u_i}_{-(i,j,k),z} + \alpha}{n^{u_i}_{-(i,j,k),\bullet} + N_T\alpha} \quad (12)$$

where $T_{-(i,j,k)}$ is the topic assignment of user rated item set $S_u$, not including the current topic $z_{i,j,k}$. $n^{item_{i,j}}_{-(i,j,k),z}$ is the number of times $item_{i,j}$ is assigned to topic $z$, $n^{u_i}_{-(i,j,k),z}$ is the number of times topic $z$ is assigned to $u_i$. $n^{\bullet}_{-(i,j,k),z}$ is the number of times topic $z$ appears, and $n^{u_i}_{-(i,j,k),\bullet}$ is the number of topics assigned to $u_i$. Both of them do not include the current instance. $N_I$ is the total number of items, while $N_T$ is the size of topic set. From these counts, we can estimate the topic-item distributions $\phi$ and user-topic distribution $\theta$ by :

$$\hat{\phi}_{item,z} = \frac{n^{item}_z + \beta}{n^{\bullet}_z + N_I\beta} \quad (13)$$

$$\hat{\theta}_{u,z} = \frac{n^u_z + \alpha}{n^{(u)}_{\bullet} + N_T\alpha} \quad (14)$$

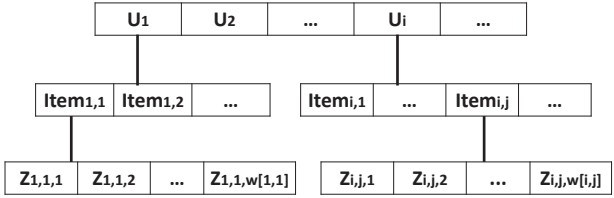

**Figure 4: Topic assignment of user documents**

---

**Algorithm 2** Algorithm for training LDA
**Input:**
  user-item rating matrix, $W = N_U \times N_I$;
**Output:**
  A set of topic assignments, $T$;
1: //initialize T;
2: Randomly assigns a topic set to each item.
3: Initialize arrays $N1, N2, N3, N4$;
4: **for** $iteration = 1$ to $l$ **do**
5:   **for** each user $i$ **do**
6:     **for** each item $j$ in $S_i$ **do**
7:       **for** each topic $k$ in $j$ **do**
8:         N1[i,k]−−;N2[i,k]−−;N3[k]−−;
9:         **for** $z = 1$ to $K$ **do**
10:
$$P[z] = \frac{(N1[j,z] + \beta) * (N2[i,z] + \alpha)}{(N3[z] + N_I * \beta) * (N4[i] + N_T * \alpha)}$$

11:         **end for**
12:         //update the topic assignment according to the P[];
13:         update(k);
14:         N1[i,k]++; N2[i,k]++; N3[k]++;
15:       **end for**
16:     **end for**
17:   **end for**
18: **end for**

---

Repeat the Gibbs sampling process to update topic assignments for several iterations, until the model parameters converge. Algorithm 2 shows the details for training LDA and inferencing the parameters, where $N1, N2, N3, N4$ are matrixes whose elements are $n^{item_{i,j}}_{-(i,j,k),z}, n^{u_i}_{-(i,j,k),z}, n^{\bullet}_{-(i,j,k),z}, n^{u_i}_{-(i,j,k),\bullet}$.

Table 1 shows two example topics that were derived from the Movielens[1] rating dataset. The table shows five movies that have the highest probability under each topic, and we can see that most movies in Topic 1 are Children's and Animation, and the movies in Topic 2 center around the genre of Action.

| Topic 1 | Topic 2 |
|---|---|
| Sleeping Beauty (1959) | Live and Let Die (1973) |
| Peter Pan (1953) | Thunderball (1965) |
| Lady and the Tramp (1955) | Lethal Weapon (1987) |
| Antz (1998) | Romancing the Stone (1984) |
| Tarzan (1999) | First Blood (1982) |

**Table 1: Two topics extracted from the Movielens dataset**

## 5. EXPERIMENTS

In this section, we conduct extensive experiments to evaluate the performance of the proposed approaches for long tail recommendation and demonstrate the superiority of our methods by comparing with other competitive techniques.

### 5.1 Experimental Setup

#### 5.1.1 State-of-the-art techniques

In this experimental study, we compare our approaches with some existing recommendation techniques for long tail recommendation on real-world data corpus collected from MovieLens and Douban[2]. Our recommendation methods have four variants, named $HT$ for Hitting Time based algorithm, $AT$ for Absorbing Time based algorithm, $AC1$ for item-based Absorbing Cost algorithm and $AC2$ for topic-based Absorbing Cost algorithm, respectively. The competitors fall in three categories which cover the state-of-the-art techniques for recommendation tasks.

- **LDA-Based Method:** Latent Dirichlet Allocation (LDA) [28] is a generative model that allows sets of observations to be explained by unobserved groups. [15, 5, 17] use LDA-based method for recommendation. Especially in [17] their empirical results suggest that association rules-based methods typically recommend rather generic, popular tags while their proposed LDA-based methods can recommend specific, infrequent tags.

- **Matrix Factorization:** Recently, several recommendation algorithms based on Matrix Factorization have been proposed [16, 2]. Authors in [6] conducted extensive experiments to evaluate the performances of various algorithms on the task of recommending long tail items, and their findings suggest that PureSVD [6] outperforms all other powerful models such as AsySVD , SVD++ [16] and classic neighborhood models.

- **Personalized PageRank:** The Personalized PageRank(PPR) [11] algorithm tends to suggest popular items since the stationary distribution for the personalized pagerank w.r.t. starting nodes $S$ is localized around the starting nodes, which combines the factors of similarity and popularity. To challenge the task of suggesting long tail items, we propose a baseline algorithm called Discounted Personalized PageRank (DPPR) which discounts items by their overall popularity. The computation of DPPR value of items w.r.t. starting nodes $S$ proceeds as follows:

$$DPPR(i|S) = \frac{PPR(i|S)}{Popularity(i)} \qquad (15)$$

where $PPR(i|S)$ denotes the personalized pagerank value of node $i$ w.r.t the starting nodes $S$, and $Popularity(i)$ is the frequency of item $i$'s rating, purchasing or downloading.

---

[1] http://www.movielens.umn.edu
[2] http://www.douban.com

### 5.1.2 Data Description

Since no benchmark data is available for performance evaluation on long tail recommendation, we utilize the real-life datasets collected from MovieLens and Douban. The long tail of a catalog is measured in terms of frequency distribution (e.g., purchases, ratings, downloads, etc.), ranked by item popularity. Though these long tail products locate in the long tail, they in the aggregate account for a sizable fraction of total consumption. We define those products as long tail products (or niche products), enjoying lowest purchasing, downloading or ratings while in the aggregate generating $r\%$ of the total sales, downloads or ratings. All the rest are called short head products. In our experiment, $r\%$ is set to $20\%$ following the 80/20 rules. We observe that about $66\%$ hard-to-find movies generate $20\%$ ratings collected by Movielens and $73\%$ least-rating books generate $20\%$ book ratings collected by Douban.

- **MovieLens Dataset:** It is a real movie dataset from the web-based recommender system MovieLens. The dataset [6] contains $1M$ ratings on a scale of 1-to-5 star for 3883 movies by 6040 users. Users rated 20-737 movies and movies received 1-583 ratings. The density for the rating matrix is $4.26\%$, a sparse matrix, which means that most users have not seen most movies.

- **Douban Dataset:** We also collect the real-life dataset from Douban for performance evaluation. Douban is a Chinese Web 2.0 website, the largest online Chinese language book, movie, and music database and one of the largest online communities in China with more than 50 million users. Users can assign 5-scale integer ratings to movies, books and music. Douban is also the largest book review website in China. We crawled totally $383,033$ unique users and $89,908$ unique books with $13,506,215$ book ratings. The density for this rating matrix is $0.039\%$, an even much sparser matrix than the Movielens.

### 5.1.3 Evaluation metrics and methodologies

As we introduced previously, long tail recommendation is a novel problem and there exists no benchmark dataset and evaluation metrics for this task. Intuitively, a desirable long tail recommender should be able to help users discover their favorite niche products and promote sales diversity, i.e., the recommended items should be in long tail (less popular), diverse, and match users' interests. According to previous study and our experience, we propose following metrics to evaluate the performance of long tail recommendation algorithms which cover various aspects of our consideration.

- **Accuracy measurement:** Our overall goal is to measure the accuracy of all mentioned algorithms and models in the long tail recommendation. As the predicted scores of recommended items by different algorithms are not in the same range, thus, to fairly compare their predictability of all algorithms, we employ the metric of Recall@N which have been widely adopted by [16, 6, 5].

- **Long tail measurement:** As mentioned in [6], to recommend popular products is easier yet more trivial. On the other hand, to recommend long tail products adds more novelty yet it is also a harder task due to data sparsity issue. In order to compare our methods with state-of-the-art techniques on the task of recommending long tail items, we evaluate them in terms of popularity. We define the popularity of recommended item as its frequency of rating.

- **Quality measurement:** Obviously, the popularity measurement is not sufficient since the recommendations may be located in the long tail but not harmony with the target user's taste. The basic idea behind all successful item-based recommender systems is to find items that are most relevant to the previous choices of the given user. Hence, to evaluate the quality of recommendation, we propose to compute the similarity between recommended items and user interests, which is similar to the method used in [19]. Besides, we conduct a user study to evaluate the usefulness of recommendations.

- **Diversity measurement:** Authors in [7] examine the effect of recommender systems on the diversity of sales, and they arrive at a surprising conclusion that most of existing recommenders can lead to reduction in sales diversity. To evaluate the recommendation diversity of all recommendation algorithms, we adopt the normalized number of different items recommended to all testing users.

- **Efficiency measurement:** This is the general performance issue of recommendation algorithms, i.e., the time cost, as the algorithms must be efficient and scalable to facilitate the online operations on large-scale datasets.

## 5.2 Performance on Recommendation

In this section, we present our performance comparison with the state-of-the-art recommendation strategies. The parameter tuning of different models is a critical issue for the system performance, such as the hyper-parameters $\alpha$ and $\beta$ for LDA, the number of latent factors (topics) for AC2, LDA and PureSVD, the dumping factor $\lambda$ for DPPR. We only report the optimal performance with tuned parameters and omit the detailed discussion due to space constraint, e.g., the default value of $\alpha$ is $50/K$ ($K$ is the number of topics), $\beta$ is $0.1$, and $\lambda$ is set as $0.5$.

### 5.2.1 Accuracy Measurement

To compare the recommendation accuracy of all algorithms in recommending long tail items, we adopt the testing methodology and the measurement Recall@N applied in [16, 6, 5]. The basic idea of this metric is to form an item set including a user favorite item and some randomly selected items, then the recommendation algorithm ranks all the items and see whether the favorite item is in the top-N results. This purely objective evaluation metric has been widely applied in recommender system evaluation.

We next introduce how this experiment will be conducted. For each dataset, known ratings are split into two subsets: training set $M$ and testing set $L$. We randomly select 4000 long tail ratings with $5$-$stars$ as the testing set and the remaining ratings as training set. As expected, the testing set is not used for training. For each long tail item $i$ rated $5$-$stars$ by user $u$ in $L$, we first randomly select 1000 additional items unrated by $u$, and then compute predicted scores for the test item $i$ as well as the additional 1000 items. Based on their scores, we form a ranked list of these 1001 items.

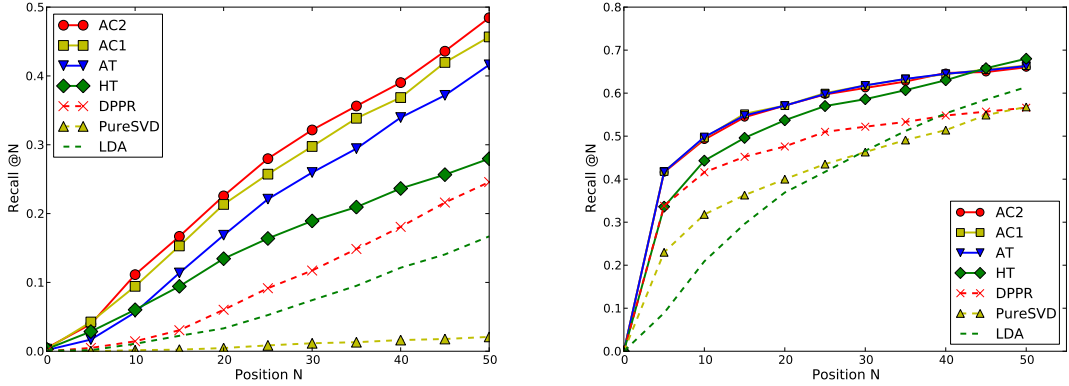

**Figure 5: (a)Recall-at-N on Movielens testset; (b)Recall-at-N on Douban testset.**

The computation of Recall@N proceeds as follows. We define hit@N for a single test case, as either the value 1 if the test item $i$ appears in the top-N results, or else the value 0. The overall Recall@N are defined by averaging all test cases:

$$Recall@N = \frac{\sum hit@N}{|L|} \qquad (16)$$

where $|L|$ is the number of all test cases, i.e., 4000 in our experiment. It should be noted that the hypothesis that all the 1000 random items are non-relevant to user $u$ tends to underestimate the computed recall with respect to true recall. However, this experimental setting and evaluation are fair to all competitors.

Figure 5(a) reports the performance of recommendation algorithms on the Movielens dataset. We only show the performance on $N$ in the range $[1...50]$, since larger value of $N$ can be ignored for a typical top-N recommendations task. It is apparent that the algorithms have significant performance disparity in terms of top-N recall. For instance, the recall of AC2 at $N = 10$ is about 0.12, and 0.48 for $N = 50$, i.e., the model has the probability of 12% to place a long tail appealing movie in top-10 and 48% in top-50. Clearly, our approaches perform better than all other competitors whose performances are less than 50% of AC2. Since the 1000 randomly selected items scatter over the whole dataset, in other words, the testing set consists of both popular and unpopular items, the existing methods generally give higher weight to the popular items in the recommendation. And hence, the target long tail item, i.e., the known user favorite, has lower probability to appear in the top-N among these 1001 items.

We also report the performances of all algorithms on Douban dataset in Figure 5(b). From the figure, we can see that the trend of comparison result is similar to that of 5(a). The main difference is that the accuracy of all algorithms increases on the Douban dataset because the Movielens dataset is smaller than Douban dataset, and the rating matrix on Movielens is much denser than that of Douban, 4.26% v.s. 0.039%. Thus, for the Movielens dataset, it is more likely that there are more relevant items to the target user in the additional 1000 items and these relevant items will have prior rankings, which degrades the recall@N performance.

From Figure 5, we can observe that AC2 outperforms the other algorithms in terms of recall, followed by AC1, AT and HT, which supports our intuitions:(1) item-based AT outperforms user-based HT since the average number of ratings per item is much higher than the average number of ratings per user, which means every item has more information to use; (2) entropy-biased AC models perform better than AT because AC can enrich the model with useful information to capture the differences of ratings from various users by utilizing the novel feature *user entropy*; (3) probabilistic topic-based entropy model beats item-based entropy model because the probabilistic topics can directly capture users' interests and tastes in the latent semantic level.

### 5.2.2 Long Tail Measurement

The above measurement Recall@N mainly evaluates the performance of all algorithms in terms of accuracy when recommending long tail items, but it has been well evidenced that being accuracy is not enough for recommender systems [21, 23, 6, 7]. To verify whether all mentioned algorithms have the desirable ability to help users accurately find the long tail items they may like in the near future, from the mixtures of both popular and unpopular items, we present the following testing methodology: given a target user $u$ and a rating log dataset, a recommendation algorithm suggests a list of items $R_u$ unrated by $u$, and we evaluate the performance of the algorithm by testing $R_u$ in measurements of Long Tail, Similarity and Diversity. We randomly sample a set of 2000 users from our datasets as the testing users. In order to speed up our proposed algorithms, we construct a subgraph centered around the query user by using breadth first search to build a candidate item set for recommendation task as illustrated in Algorithm 1. In this study, the default size of candidate item set, $\mu$, is $6k$ and the default iterations $\tau$ is 15. In the following, we first present the performance of all algorithms in the measurement of Long Tail.

In this experiment, we evaluate the popularity of the recommended items over the 2000 testing users, and report the average numbers of ratings. The comparison results are shown in Figure 6(a) and 6(b). From the comparisons, we can observe that no matter how many recommendations are returned, our proposed recommendation algorithms consistently recommend more niche items than other existing methods, which shows the effectiveness of our methods in recommending long tail products. As expected, our designed baseline algorithm DPPR shows comparable performance with our approaches, as it discounts items by their popularity and gives priority to long tail items. However, it performs worse in terms of Recall@N as shown in previous analysis. An interesting observation is that the popular items enjoy priority to be recommended in LDA and PureSVD models, so the top suggested results from the two latent factor-based recommendation models are more

904

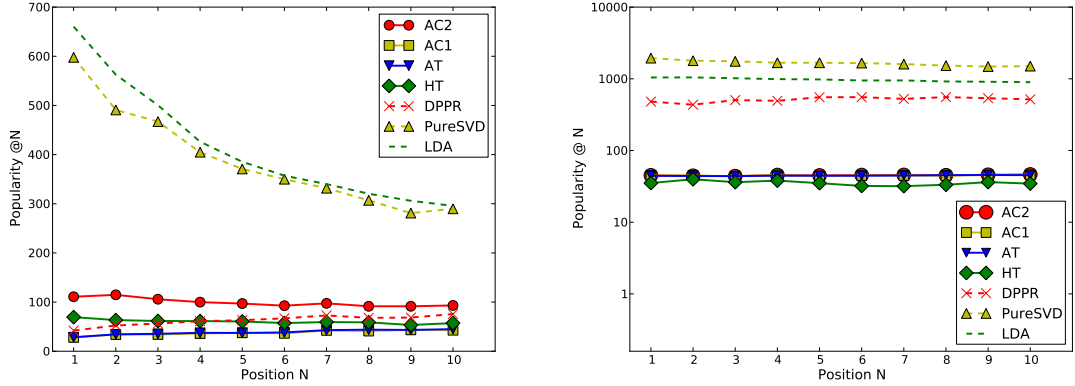

Figure 6: (a)Popularity at N on Douban testset; (b)Popularity at N on Movielens testset.

likely to be popular items. That is why Popularity@N decreases with the increasing $N$ for LDA and PureSVD in Figure 6. It should be noted that the decrease of Popularity@N is not visibly obvious in Figure 6(b) since we take the logarithm form on the vertical axis for the Movielens testset.

### 5.2.3 Diversity Measurement

As we discussed previously, most of existing recommenders can lead to a reduction in sales diversity. Because these recommenders suggest products based on sales and ratings, they may create a rich-get-richer effect for popular products. Some recommenders can push users to new products, but they often push different users toward the same products. Thus the overall recommendation diversity will be degenerated rather than enhanced. To evaluate the recommendation diversity of all recommendation algorithms, we propose the computation of diversity as follows:

$$Diversity = \frac{|\bigcup_{u \in U} R_u|}{|I|} \quad (17)$$

where $U$ is the set of users and $I$ is the set of items. $R_u$ denotes the set of recommended items for $u$.

**Table 2: Comparison on Diversity**

|  | AC2 | AC1 | AT | HT | DPPR | PSVD | LDA |
|---|---|---|---|---|---|---|---|
| Douban | 0.58 | **0.625** | 0.58 | 0.55 | 0.45 | 0.325 | 0.035 |
| Movielens | 0.42 | **0.425** | 0.42 | 0.41 | 0.35 | 0.245 | 0.025 |

In our experiment, for Douban dataset $U$ consists of 2000 testing users, and an ideal recommender model can recommend 20000 ($|I| = 20000$) unique items at most if each testing user is suggested a list of 10 items. As Table 2 shows, our proposed recommenders outperform the competitors in terms of Diversity. Among all algorithms, LDA performs worst. Though LDA can recommend user-interested items for each user, it only recommends about 700 different items for the 2000 testing users on Douban dataset. Among our approaches, AC1 performs best and it can suggest 12500 unique items for the 2000 testing users; AT performs as well as AC2 and both of them can recommend 11600 unique items. The experimental results also confirm that item-based methods(e.g., AT algorithm) perform better than user-based methods(e.g., HT algorithm) in terms of recommendation diversity: HT suggests 11000 unique items while AT recommends 11600 unique items for the 2000 testing users. In Movielens dataset, the trend of comparison result is similar to the result in Douban dataset. The main difference is that the Diversity of almost all algorithms decreases in Movielens dataset due to that the rating matrix on Movielens is much denser than that of Douban. Two randomly selected users in Movielens dataset enjoy higher probability of having rated same items and enjoying the similar interests than two randomly selected users in Douban, which increases the probability that the same items are suggested for different users. Based on the above observations, we can conclude that our proposed algorithms can help firms to improve their sales diversity by recommending diverse items to users.

### 5.2.4 Similarity Measurement

Obviously, both popularity and diversity measurements are not sufficient since the recommendations may reside in the long tail but not match the interest of target user. In this part, we evaluate the performance of all algorithms in similarity measurement on Douban dataset.

We utilize a well-defined book ontology from dangdang[3], an Amazon-like Chinese e-commerce company, and map the douban's books into this ontology by books' ISBN. The computation of similarity in this ontology proceeds as follows. Given a book, we first find category matches in the form of paths between categories. For instance, given the book "Introduction to Data Mining", we would identify the hierarchial category "Book: Computer & Internet: Database: Data Mining and Data Warehouse: Introduction to Data Mining" while the hierarchial category for book "Information Storage and Management" is "Book: Computer & Internet: Database: Data Management: Information Storage and Management", where ":" is used to separate different categories. Hence, to measure how similar two books are, we can use a notion of similarity between the corresponding categories provided by the well defined book ontology. In particular, we measure the similarity between two categories $C_i$ and $C_j$ as the length of their longest common prefix $P(C_i, C_j)$ divided by the length of longest path between $C_i$ and $C_j$. More precisely, the similarity is defined as:

$$Sim(C_i, C_j) = \frac{|P(C_i, C_j)|}{\max(|C_i|, |C_j|)} \quad (18)$$

where $|C_i|$ denotes the length of path $C_i$. For instance, the similarity between the above two books is 2/4 since they share the path "Book: Computer & Internet: Database" and the longest path is 4. We evaluate the similarity between two books by measuring the similarity between their corresponding categories.

---
[3] http://book.dangdang.com/

Based on the above equation, we propose a novel measurement to measure the relevance between a user $u$ and an item $i$, as follows

$$Sim(u,i) = \max_{j \in S_u} sim(i,j) \quad (19)$$

where $S_u$ is $u$'s preferred item set. The definition in Eq.(19) indicates that: the recommended book $i$ is relevant to the target user's taste if book $i$ is similar with one of his favorite books. We evaluate the relevance of the recommendation results of 2000 testing users and we report the average values.

Table 3: Comparison on Similarity

| AC2 | AC1 | AT | HT | DPPR | PureSVD | LDA |
|---|---|---|---|---|---|---|
| **0.48** | 0.42 | 0.39 | 0.37 | 0.36 | 0.45 | 0.43 |

The comparison results are shown in Table 3. From the table, we can see that our proposed recommendation methods can perform as well as, if not better than, other state-of-the-art recommendation algorithms. Especially, our proposed AC2 variant performs best among all algorithms, which supports our intuition that various user-item rating pairs should be treated differently in the recommendation. Among all of our proposed algorithms, we can see that item-based AT, AC1 and AC2 can recommend more relevant items than user-based HT, and probabilistic topic-based AC2 can suggest more similar items than item-based AC1 due to that the topic-based *user entropy* is more proximate to the exact *user entropy* than the item-based. Our designed baseline algorithm DPPR does not perform well on the measurement of similarity although it can recommend long tail items.

### 5.2.5 Impact of Parameter $\mu$

As our approach is an item-oriented solution, we have to decide a proper candidate set for recommendation to avoid time-consuming dataset scan. In this experiment, we investigate how the number of candidate items in the constructed subgraph $\mu$ will affect the effectiveness and efficiency of the recommendation. Note that, processing on the whole graph can be time-consuming and inefficient as analyzed previously. In practice, we can adopt a subgraph with fewer candidate item nodes to make recommendations, which can achieve almost the same performance as the the whole graph scan, as shown in Table 4. Due to the space constraint, we only show the performance of AC2 on Douban dataset by varying $\mu$ in the table.

Table 4: Impact of Parameter $\mu$

| $\mu$ | 3000 | 4000 | 5000 | 6000 | 89908 |
|---|---|---|---|---|---|
| Popularity | 100.6 | 100.1 | 95.7 | **93.2** | 94.8 |
| Similarity | 0.44 | 0.46 | 0.47 | **0.48** | **0.48** |
| Diversity | **0.585** | **0.585** | 0.58 | 0.58 | 0.58 |
| Efficiency(s) | **0.17** | 0.3 | 0.42 | 0.52 | 12.7 |

From Table 4, we can observe that (1) Popularity slightly decreases with the increasing $\mu$ ; (2) Similarity increases with $\mu$ increasing from $3k$ to $6k$ and then does not change much when $\mu$ is larger than $6k$; (3) Diversity slightly decreases with the increasing $\mu$ because an item would enjoy higher probability to appear in more candidate sets with the increasing $\mu$, which may result in the decrease of recommendation diversity; (4) the time cost of AC2 algorithm increases with parameter $\mu$ increasing from 3000 to 89908. As shown in Table 4, AC2 can achieve satisfactory performance with relatively small $\mu$, e.g., $3k$-$6k$, and the performance does not change much when $\mu$ is larger than $6k$. Based on the above analysis, we can conclude that our proposed methods are scalable to large datasets by selecting much smaller subgraph centered around the target user, which ensures the efficiency of our algorithms.

### 5.2.6 Comparison on Efficiency

We next proceed to perform an efficiency comparison of different recommendation algorithms. All the recommendation algorithms are implemented in Java (JDK 1.6) and run on a Linux Server with 32G RAM. They are required to recommend 10 items for each user on Douban dataset. Since LDA and PureSVD are model based methods, we only report their online recommendation time costs, ignoring their offline training time. For our proposed AC2 algorithm, we present its time cost with parameter $\mu = 6k$ and $\tau = 15$, which also excludes the offline time cost of learning *entropy of users* with LDA model.

Table 5: Comparison on Time Cost

|  | LDA | PureSVD | AC2 | DPPR |
|---|---|---|---|---|
| time(s) | 0.47 | 0.45 | 0.52 | 13.5 |

From the Table 5, we can observe that our proposed AC2 algorithm achieves comparable performance with model-based approaches such as LDA and PureSVD, and it beats the graph-based DPPR algorithm on Douban dataset. Our approach only needs to explore the relevant subgraph to avoid time-consuming global graph scan, while preserving the effectiveness of the recommendation. Note that, though the latent factor model based approaches, such as LDA and PureSVD, are quite efficient in our study, their costs are linear to the size of data. For the latent factor model, we have to compute the similarity of a user to all the items in the dataset and subsequently select the best $k$ among them to recommend to the user. When the dataset becomes larger (e.g., containing millions of items), computing the top-k items for each user requires millions of vector operations per user.

### 5.2.7 User Study

In order to evaluate the usefulness of our proposed recommendation algorithms, we employ a user survey on Movielens dataset by hiring 50 movie-lovers as evaluators to answer the evaluation questions. In the survey, we use AC2, DPPR, PureSVD and LDA algorithms to recommend 10 movies for each evaluator according to their preferences, respectively.

The survey is structured as follows. In order to elicit the evaluators' preferences, each one is first presented 50 movies and required to provide their ratings. It should be noted that these 50 movies almost cover all movie genres. They are then offered 10 recommendations, represented as a list of ten movies titles (together with the corresponding DVD covers and IMDB links) and asked to answer the following evaluation questions for each movie:

- Preference: How much does the recommended movie match your taste and interest (1-5)?

- Novelty: Have you ever known the recommended movie before (0(Y) - 1(N))? If yes, how did you know it(free text)?

- Serendipity: How much surprise and serendipity can the movie bring to you (1-5)?

- Score : Please provide the overall rating for the movie (1-5).

The 50 evaluators individually answer the above evaluation questions and the average results are given in Table 6. Based on this user study, we have following observations: (1) Our proposed method can recommend more satisfactory movies to users in overall. Specifically, the movies suggested by our method not only match users'

906

tastes and interests, but also bring more novelty and serendipity to users due to that the movies suggested by our method are much more likely to lie in the long tail and hence most users may have never known them before. (2) Compared with our recommendations, the movies recommended by LDA and PureSVD concentrate on the short head, the majority of which are hit movies. According to the evaluators' answers and explanations to *Novelty*, we observe that they have known more than one-third of the recommended movies before, from other medium such as film posters, newspapers, broadcast and their friends. Besides, some evaluators explained that they have seen the movies in the top lists of some websites such as IMDB, Hulu and YouTube. Obviously, this kind of recommendation may not be that useful. (3) Although the movies recommended by DPPR can also bring novelty to users since most of them are in the long tail, they do not match users' tastes and interests very well. That is why the movies suggested by DPPR have lower overall scores.

Table 6: Comparison on Usefulness

|         | Preference | Novelty | Serendipity | Score |
|---------|------------|---------|-------------|-------|
| AC2     | 4.32       | **0.98**| 4.78        | **4.41** |
| DPPR    | 3.12       | 0.89    | 3.95        | 3.65  |
| PureSVD | **4.34**   | 0.64    | 2.12        | 4.25  |
| LDA     | 4.12       | 0.66    | 2.15        | 4.22  |

## 6. CONCLUSIONS

In this paper, we have addressed the problem of long tail recommendation which aims to suggest niche items to users. We first analyzed the long tail phenomenon and long tail recommendation problem. Based on the undirect edge-weighted graph representation, four recommendation algorithm variants were proposed which utilized hitting time, absorbing time and absorbing cost. Our approaches can exploit the less popular items residing in the long tail of inventory and emphasize the user interests and recommendation diversity. We conducted extensive experiments on two real datasets and the experimental results show that our proposed algorithms outperform the state-of-the-art recommendation algorithms for long tail item recommendation task in terms of recommendation accuracy, quality and diversity. Our work can serve as an alternative of recommender system and provide a potential and novel feature for online sale services.

## 7. ACKNOWLEDGMENTS

This research was supported by the National Natural Science Foundation of China under Grant No. 60933004 and 61073019.

## 8. REFERENCES

[1] D. Agarwal and B.-C. Chen. Regression-based latent factor models. In *KDD*, pages 19–28, 2009.
[2] D. Agarwal and B.-C. Chen. flda: matrix factorization through latent dirichlet allocation. In *WSDM*, pages 91–100, 2010.
[3] C. Anderson. *The Long Tail: Why the Futhure of Business is Selling Less of More*. Hyperion, 2006.
[4] M. Brand. A random walks perspective on maximizing satisfaction and profit. In *SDM*, 2005.
[5] W.-Y. Chen, J.-C. Chu, J. Luan, H. Bai, Y. Wang, and E. Y. Chang. Collaborative filtering for orkut communities: discovery of user latent behavior. In *WWW*, pages 681–690, 2009.
[6] P. Cremonesi, Y. Koren, and R. Turrin. Performance of recommender algorithms on top-n recommendation tasks. In *RecSys*, pages 39–46, 2010.
[7] D. M. Fleder and K. Hosanagar. Recommender systems and their impact on sales diversity. In *EC*, pages 192–199, 2007.
[8] F. Fouss, A. Pirotte, J. Renders, and M. Saerens. A novel way of computing dissimilarities between nodes of a graph, with application to collaborative filtering and subspace projection of the graph nodes. In *ECML*, 2004.
[9] T. L. Griffiths and M. Steyvers. Finding scientific topics. *Proc Natl Acad Sci U S A*, pages 5228–35, 2004.
[10] Z. Guan, J. Wu, Q. Zhang, A. Singh, and X. Yan. Assessing and ranking structural correlations in graphs. In *SIGMOD*, pages 937–948, 2011.
[11] T. H. Haveliwala. Topic-sensitive pagerank. In *WWW*, pages 517–526, 2002.
[12] J. Herlocker, J. A. Konstan, and J. Riedl. An empirical analysis of design choices in neighborhood-based collaborative filtering algorithms. *Inf. Retr.*, 5:287–310, October 2002.
[13] J.G.Kemeny and J.L.Snell. *Finite Markov Chains*. Springer-Verlag, 1976.
[14] J.Norris. *Markov Chains*. Cambridge University Press, 1997.
[15] Y. Kang and N. Yu. Soft-constraint based online lda for community recommendation. In *PCM*, pages 494–505, 2010.
[16] Y. Koren. Factorization meets the neighborhood: a multifaceted collaborative filtering model. In *KDD*, pages 426–434, 2008.
[17] R. Krestel, P. Fankhauser, and W. Nejdl. Latent dirichlet allocation for tag recommendation. In *RecSys*, pages 61–68, 2009.
[18] L. Lovsz. Random walks on graphs: A survey, 1993.
[19] H. Ma, M. R. Lyu, and I. King. Diversifying query suggestion results. In *AAAI*, 2010.
[20] M. R. McLaughlin and J. L. Herlocker. A collaborative filtering algorithm and evaluation metric that accurately model the user experience. In *SIGIR*, pages 329–336, 2004.
[21] S. M. McNee, J. Riedl, and J. A. Konstan. Being accurate is not enough: how accuracy metrics have hurt recommender systems. In *CHI EA*, pages 1097–1101, 2006.
[22] D. Oktar. *Recommendation Systems: Increasing Profit by Long Tail*. http://en.webrazzi.com/2009/09/18/.
[23] K. Onuma, H. Tong, and C. Faloutsos. Tangent: a novel, 'surprise me', recommendation algorithm. In *KDD*, pages 657–666, 2009.
[24] A. G. Parameswaran, G. Koutrika, B. Bercovitz, and H. Garcia-Molina. Recsplorer: recommendation algorithms based on precedence mining. In *SIGMOD*, pages 87–98, 2010.
[25] F. Ricci, L. Rokach, B. Shapira, and P. B. Kantor. *Recommender Systems Handbook*. Springer-Verlag New York, Inc., 2010.
[26] C. E. Shannon. Prediction and entropy of printed english. *Bell Systems Technical Journal*, 30:50–64, 1951.
[27] S.Ross. Stochastic processes,2nd ed, 1996.
[28] M. Steyvers and T. Griffiths. *Probabilistic Topic Models*. Lawrence Erlbaum Associates, 2007.
[29] F. Wang, S. Ma, L. Yang, and T. Li. Recommendation on item graphs. In *ICDM*, pages 1119–1123, 2006.